\documentclass[aps,prl,reprint,amsmath,amssymb,superscriptaddress,nofootinbib]{revtex4-2}

\usepackage{graphicx}
\usepackage{hyperref}
\usepackage{braket}
\usepackage{mathtools}
\usepackage{empheq}
\usepackage{xcolor}

\begin{document}

\title{Non-Perturbative Closed Form for the Typical Bipartite\\
Mutual Information of Haar-Random States}

\author{Zhi-Wei Wang}
\email{zhiweiwang.phy@gmail.com}
\affiliation{College of Physics, Jilin University,
Changchun, 130012, People's Republic of China}
\affiliation{Computer Science, University of York, York YO10 5GH, United Kingdom}

\author{Pei-Wen Li}
\affiliation{College of Physics, Jilin University,
Changchun, 130012, People's Republic of China}

\author{Samuel L.\ Braunstein}
\email{sam.braunstein@york.ac.uk}
\affiliation{Computer Science, University of York, York YO10 5GH, United Kingdom}

\date{\today}

\begin{abstract}
The average bipartite quantum mutual information
$\langle I(A{:}B)\rangle$ of Haar-random pure states can be
expressed exactly through Page's formula in terms of digamma
functions.  We show that this quantity admits a single
non-perturbative closed form:
$\langle I(A{:}B)\rangle =
(d_A^2-1)(d_B^2-1)\,\mathcal{G}(d_A,d_B,d_E)$, where
$\mathcal{G}$ is given by an explicit convergent integral over a
Bose--Einstein kernel.  The overall factor
$(d_A^2-1)(d_B^2-1)=\dim[\mathfrak{su}(d_A)]\cdot\dim[\mathfrak{su}(d_B)]$
is exact, not merely asymptotic.
The asymptotic expansion of $\mathcal{G}$ in $1/N$ yields a
Bernoulli-factorised series whose coefficients involve
$\zeta(1{-}2k)$; this series diverges, and our integral is its exact
Borel sum.
The integral representation also makes
$\langle I\rangle < (d_A^2{-}1)(d_B^2{-}1)/(2N)$ manifest via a
scale-inversion symmetry of the kernel.
Our derivation traces the
mutual information's structure to an exact decomposition of Page's
entropy into a diagonal (Dirichlet) contribution and a
Schur-majorisation eigenvalue correction, whose assembly into the
mutual information cleanly separates classical from quantum
correlations.
\end{abstract}

\maketitle

\emph{Introduction.}---A foundational result in quantum information
is that small subsystems of large quantum systems generically
appear thermal~\cite{Goldstein2006,Popescu2006,DAlessio2016}.
The quantitative backbone is Page's formula~\cite{Page1993,Foong1994,Sen1996}
for the average von Neumann entropy of a random pure state, which
underpins typicality arguments, the eigenstate thermalisation
hypothesis (ETH), and information-theoretic treatments of black hole
evaporation~\cite{Hayden2007}.  The leading behaviour, finite-size
corrections, and fluctuation statistics of individual subsystem
entropies are all well characterised~\cite{Bianchi2019}, but the
structure of the bipartite mutual information
$I(A{:}B) \equiv S(A)+S(B)-S(AB)$, which captures the total
correlations between two subsystems embedded in a larger Hilbert
space, has received less attention.  In finite-dimensional systems
such as Noisy Intermediate-Scale Quantum (NISQ)
devices~\cite{Preskill2018}, mesoscopic quantum heat engines, and
small-subsystem models of black hole information~\cite{Hayden2007},
the typical mutual information controls the strength of bipartite
correlations that survive partial tracing, and its finite-size
corrections determine how quickly the thermodynamic limit is
approached.  In principle the typical value
$\langle I(A{:}B)\rangle$ is fully determined by Page's formula,
but the resulting expression is a combination of digamma functions
and rational corrections from which structural properties are not
immediately visible.  Sen's Bernoulli expansion~\cite{Sen1996} of
individual subsystem entropies is well known, and asymptotic analyses
of R\'enyi entropies and their variances have been developed in the
random-matrix literature~\cite{Bianchi2019}; however, the structural
properties of the \emph{mutual information} (in particular its
factorisation, its non-perturbative completion, and its dimensional
bounds) have not been systematically examined.

In this Letter we show that a single integral representation makes
all structural properties of $\langle I(A{:}B)\rangle$ transparent.
Our main results are:
(i)~the exact typical mutual information factors as
$(d_A^2{-}1)(d_B^2{-}1)$ times a dimension-dependent function
$\mathcal{G}$, making the vanishing of correlations for trivial
subsystems ($d=1$) manifest at every order, not merely asymptotically;
(ii)~$\mathcal{G}$ is given by a convergent Bose--Einstein integral
that is the exact Borel sum of the divergent Bernoulli asymptotic
series in $1/N$; and
(iii)~a scale-inversion symmetry of the integrand makes the strict
upper bound
$\langle I\rangle < (d_A^2{-}1)(d_B^2{-}1)/(2N)$
manifest by folding the integral onto a finite domain where the
integrand is pointwise positive.

We consider a pure state
$\ket{\Psi}\in\mathcal{H}_A\otimes\mathcal{H}_B\otimes\mathcal{H}_E$,
drawn from the Haar measure, with subsystem dimensions $d_A$, $d_B$,
$d_E$ and total dimension $N=d_Ad_Bd_E$.  Throughout, we assume
$d_Ad_B\le d_E$, which ensures Page's formula evaluates $S(AB)$
without swapping subsystem and environment dimensions.The underlying geometric framework---including exact finite-size distributions from the hyperspherical projected central limit theorem, and the Lie-algebraic variance structure---is detailed in a simultaneously submitted companion paper \cite{Wang2026PRA}.

\emph{Diagonal-entropy decomposition.}---The starting point is
Page's formula for the average entropy of a subsystem of dimension
$m$ in an environment of dimension $n$~\cite{Page1993,Foong1994,Sen1996}:
\begin{equation}
\langle S_{\mathrm{vN}}\rangle
= \psi(mn{+}1) - \psi(n{+}1) - \frac{m-1}{2n}\,,
\label{eq:page}
\end{equation}
where $\psi$ denotes the digamma function.  For a tripartite system
with $N = d_Ad_Bd_E$, the three entropies entering $\langle I\rangle$
use $(m,n) = (d_A,\, d_Bd_E)$, $(d_B,\, d_Ad_E)$, and
$(d_Ad_B,\, d_E)$ respectively; the condition $d_Ad_B\le d_E$ ensures
the last of these does not swap subsystem and environment.

We observe that Eq.~(\ref{eq:page}) splits naturally into two pieces
with distinct physical origins.

The digamma piece is the average \emph{diagonal entropy}%
~\cite{Lloyd1988,Wootters1990},
\begin{equation}
\langle S_{\mathrm{diag}}\rangle
\equiv \Bigl\langle -\!\sum_k P_k \ln P_k\Bigr\rangle
= \psi(mn{+}1)-\psi(n{+}1),
\label{eq:Sdiag}
\end{equation}
defined as the Shannon entropy of the diagonal elements
$P_k=\rho_{kk}$ in a fixed basis.
This identity follows from the fact that Haar-random states induce a
symmetric Dirichlet distribution
$\{P_k\}\sim\mathrm{Dir}(n,\ldots,n)$ on the probability
simplex~\cite{Lloyd1988,Zyczkowski2001,Bengtsson2006}: for each
component,
$\langle P_k\ln P_k\rangle = \tfrac{1}{m}[\psi(n{+}1)-\psi(mn{+}1)]$,
and summing over $k$ yields Eq.~(\ref{eq:Sdiag}).

The remaining piece, $-(m{-}1)/(2n)$, is the average deficit
between diagonal and eigenvalue entropies.  By Schur's majorisation
theorem~\cite{Schur1923,Marshall2011}, the eigenvalue vector of a
Hermitian matrix majorises its diagonal, so
$S_{\mathrm{vN}}\le S_{\mathrm{diag}}$~\cite{Wehrl1978}.  The
average gap is exactly $(m{-}1)/(2n)$, where the numerator $(m{-}1)$
is the dimension of the probability simplex.  This is also the
dimension of the Cartan subalgebra of $\mathfrak{su}(m)$%
~\cite{Georgi1999}: the traceless diagonal generators span the
independent directions on the simplex constrained by
$\sum_k P_k = 1$.

Applying this split to each entropy in
$\langle I(A{:}B)\rangle = \langle S(A)\rangle + \langle S(B)\rangle
- \langle S(AB)\rangle$ gives
\begin{equation}
\langle I(A{:}B)\rangle
= \langle I_{\mathrm{diag}}\rangle + \Delta_{\mathrm{ev}}\,,
\label{eq:Isplit}
\end{equation}
where the diagonal mutual information is
\begin{eqnarray}
\langle I_{\mathrm{diag}}\rangle
&=& \psi(N{+}1)-\psi\!\bigl(\tfrac{N}{d_A}{+}1\bigr)
 -\psi\!\bigl(\tfrac{N}{d_B}{+}1\bigr)\nonumber\\
 &&+\,\psi\!\bigl(\tfrac{N}{d_Ad_B}{+}1\bigr),
\label{eq:Idiag}
\end{eqnarray}
and the eigenvalue correction assembles, after straightforward
algebra, to
\begin{eqnarray}
\Delta_{\mathrm{ev}}
&=& -\frac{d_A(d_A{-}1)}{2N}
   -\frac{d_B(d_B{-}1)}{2N}
   +\frac{d_Ad_B(d_Ad_B{-}1)}{2N} \nonumber\\
&=& \frac{(d_A^2{-}1)(d_B^2{-}1)-(d_A{-}1)(d_B{-}1)}{2N}\,.
\label{eq:Deltaev}
\end{eqnarray}
The first term in $\Delta_{\mathrm{ev}}$ counts
$\dim[\mathfrak{su}(d_A)\otimes\mathfrak{su}(d_B)]
=(d_A^2{-}1)(d_B^2{-}1)$; the subtracted term counts
$\dim[\mathrm{Cartan}(d_A)\otimes\mathrm{Cartan}(d_B)]
=(d_A{-}1)(d_B{-}1)$.  The physical interpretation is as follows.
The diagonal mutual information
$\langle I_{\mathrm{diag}}\rangle$ captures correlations
between the \emph{classical probability distributions} on the
respective simplices of $\rho_A$, $\rho_B$, and $\rho_{AB}$: it is
the mutual information one would compute from the diagonal elements
alone, without access to off-diagonal coherences.  The eigenvalue
correction $\Delta_{\mathrm{ev}}$ captures the additional mutual
information arising because the von Neumann entropy uses eigenvalues
rather than diagonal elements.  By Schur's majorisation theorem, this
correction is strictly positive: eigenvalue entropy is bounded above
by diagonal entropy for each subsystem individually, but the
subtraction $S(A)+S(B)-S(AB)$ flips the sign for the composite $AB$
term, producing a net positive quantum-coherence contribution that
dominates by a factor of $\sim(d_Ad_B+d_A+d_B)$ over the classical
piece in the large-dimension regime.
At leading order in $1/N$,
$\langle I_{\mathrm{diag}}\rangle\sim (d_A{-}1)(d_B{-}1)/(2N)$
and the total is $(d_A^2{-}1)(d_B^2{-}1)/(2N)$.

We emphasise that this Cartan/off-diagonal separation enters through
the \emph{entropy}, not through the purity.  {For a subsystem of dimension $m$ and environment $n$ (with total dimension $mn$),} in the generalised Bloch
decomposition
$\rho = \mathbb{I}/{m} + \frac{1}{2}\sum_a r_a\lambda_a$, where
$\{\lambda_a\}$ are the {$m^2{-}1$} generators of $\mathfrak{su}({m})$
normalised so that
$\mathrm{Tr}(\lambda_a\lambda_b)=2\delta_{ab}$~\cite{Kimura2003},
the purity decomposes as
$\mathrm{Tr}(\rho^2) = 1/{m} + \frac{1}{2}\sum_a r_a^2$.
The Cartan generators are diagonal, so the diagonal purity
$\sum_k P_k^2 = 1/{m} + \frac{1}{2}\sum_{a\in\mathrm{Cartan}} r_a^2$.
Combining Lubkin's
formula~\cite{Lubkin1978}
$\langle\mathrm{Tr}(\rho^2)\rangle = ({m{+}n})/({mn}{+}1)$ with the
diagonal second moment
$\sum_k\langle P_k^2\rangle = ({n}{+}1)/({mn}{+}1)$ yields,
after subtracting $1/{m}$, the total Cartan and off-diagonal variances.
Dividing by the respective generator counts $({m}{-}1)$ and {$m(m{-}1)$}
gives
\begin{equation}
\langle r_a^2\rangle_{\mathrm{Cartan}}
= \langle r_a^2\rangle_{\mathrm{off\text{-}diag}}
= \frac{2}{{m(mn{+}1)}}\,.
\label{eq:equalvar}
\end{equation}
The Haar measure distributes variance
\emph{democratically} across all {$m^2{-}1$} generators. This is a direct 
geometric consequence of the unitary invariance of the ensemble, 
mathematically mirroring the uniform variance scaling that emerges when 
exactly projecting high-dimensional hyperspheres onto lower-dimensional 
subspaces~\cite{Wang2023,Wang2024}. The
separation of Cartan from off-diagonal contributions arises only
when the nonlinear $-x\ln x$ of the von Neumann entropy
encounters the probability-simplex boundary, precisely through
the $-(m{-}1)/(2n)$ correction in Page's formula.

\emph{Bernoulli-factorised asymptotic expansion.}---Applying the
digamma asymptotics
$\psi(z{+}1)\sim\ln z + 1/(2z) - \sum_{k=1}^\infty B_{2k}/(2kz^{2k})$
to each term in Eq.~(\ref{eq:Idiag}), and using the algebraic identity
$(d_A^{2k}{-}1)+(d_B^{2k}{-}1)-(d_A^{2k}d_B^{2k}{-}1)
= -(d_A^{2k}{-}1)(d_B^{2k}{-}1)$,
the full expansion collapses to
\begin{equation}
\langle I\rangle
\sim \frac{(d_A^2{-}1)(d_B^2{-}1)}{2N}
-\!\sum_{k=1}^{\infty}\frac{B_{2k}}{2kN^{2k}}
(d_A^{2k}{-}1)(d_B^{2k}{-}1).
\label{eq:bernoulli}
\end{equation}
Every order $k\ge 1$ carries the symmetric factor
$(d_A^{2k}{-}1)(d_B^{2k}{-}1)$; no mixed terms such as
$d_A^2 d_B^4$ appear, all odd inverse powers of $N$ beyond
the leading term vanish identically, and successive corrections
alternate in sign.  Writing out the first few orders explicitly:
\begin{eqnarray}
\langle I\rangle
&\sim& \frac{(d_A^2{-}1)(d_B^2{-}1)}{2N}
     -\frac{(d_A^2{-}1)(d_B^2{-}1)}{12N^2}\nonumber\\
   && +\frac{(d_A^4{-}1)(d_B^4{-}1)}{120N^4}
     -\frac{(d_A^6{-}1)(d_B^6{-}1)}{252N^6}+\cdots\,.
\label{eq:explicit}
\end{eqnarray}
The first two orders share the prefactor $(d_A^2{-}1)(d_B^2{-}1)$
before higher Casimir-like factors enter at $\mathcal{O}(1/N^4)$.
The absence of odd inverse powers beyond the leading $\mathcal{O}(1/N)$
term is a direct consequence of the digamma asymptotics: the
Bernoulli numbers $B_{2k+1}=0$ for $k\ge 1$.  If either subsystem
is trivial ($d_A=1$ or $d_B=1$), every term in the expansion
vanishes: the factorised form makes this transparent without
requiring separate analysis.

The identity $-B_{2k}/(2k)=\zeta(1{-}2k)$ allows
Eq.~(\ref{eq:bernoulli}) to be rewritten as
\begin{equation}
\langle I\rangle
\sim \frac{(d_A^2{-}1)(d_B^2{-}1)}{2N}
+\sum_{k=1}^{\infty}\frac{\zeta(1{-}2k)}{N^{2k}}
(d_A^{2k}{-}1)(d_B^{2k}{-}1),
\label{eq:zeta}
\end{equation}
placing the finite-size corrections in structural parallel with
zeta-regularised vacuum sums (e.g., $\zeta(-1)=-1/12$ for the
bosonic string).  Crucially, since
$|B_{2k}|\sim 2(2k)!/(2\pi)^{2k}$ the series \emph{diverges}:
the Bernoulli expansion is asymptotic, not convergent.  Truncating
at any finite order cannot recover $\langle I\rangle$ to arbitrary
precision.

\emph{Non-perturbative closed form.}---We now derive a single
convergent integral that exactly Borel-sums the divergent
series~(\ref{eq:bernoulli}).  The key tool is Binet's second formula
for the digamma function~\cite{NIST}:
\begin{equation}
\psi(z{+}1) = \ln z + \frac{1}{2z}
- 2\int_0^\infty \frac{t\,dt}{(t^2{+}z^2)(e^{2\pi t}{-}1)}\,.
\label{eq:binet}
\end{equation}
Applying this to the four digamma terms in
Eq.~(\ref{eq:Idiag}) with
$z = N,\, N/d_A,\, N/d_B,\, N/(d_Ad_B)$ respectively, the logarithmic
pieces cancel pairwise
($\ln N - \ln(N/d_A)-\ln(N/d_B)+\ln(N/d_Ad_B)=0$).  The
$1/(2z)$ pieces assemble to
$\frac{1}{2}\bigl[\frac{1}{N}-\frac{d_A}{N}-\frac{d_B}{N}
+\frac{d_Ad_B}{N}\bigr] = (d_A{-}1)(d_B{-}1)/(2N)$,
which combines with
$\Delta_{\mathrm{ev}}$ from Eq.~(\ref{eq:Deltaev}) to give a
total rational piece of $(d_A^2{-}1)(d_B^2{-}1)/(2N)$.  The
remaining Binet integrals, after the substitution $u=t/d_E$
(so that $t^2+z^2 \to d_E^2(u^2+\alpha^2)$ with
{$\alpha = z/d_E$}, giving poles at
$\alpha = d_Ad_B,\, d_B,\, d_A,\, 1$), combine into a single
integral:
\begin{widetext}
\begin{equation}
\boxed{
\begin{aligned}
\langle I(A{:}B)\rangle &= (d_A^2-1)(d_B^2-1)\;\Bigl[\frac{1}{2N}
-2\,\mathcal{J}(d_A,d_B,d_E)\Bigr],\\[6pt]
\mathcal{J}(d_A,d_B,d_E) &\equiv
\int_0^\infty
\frac{u\,(d_A^2 d_B^2-u^4)\,du}
{(e^{2\pi ud_E}-1)\,
(u^2{+}1)(u^2{+}d_A^2)(u^2{+}d_B^2)(u^2{+}d_A^2d_B^2)}\,.
\end{aligned}
}
\label{eq:closed}
\end{equation}
\end{widetext}
The extraction of $(d_A^2{-}1)(d_B^2{-}1)$ from the integral is
exact.  To see why, decompose the rational kernel via partial
fractions in $w=u^2$:
\begin{equation}
\frac{u(C^2{-}u^4)}{\prod_i(u^2{+}\alpha_i^2)}
= \frac{1}{(d_A^2{-}1)(d_B^2{-}1)}
\sum_i \frac{s_i\, u}{u^2{+}\alpha_i^2}\,,
\label{eq:pf}
\end{equation}
where $(\alpha_i) = (1,d_A,d_B,C)$, $C=d_Ad_B$, and
$(s_i) = (+1,-1,-1,+1)$.  All four residues share the magnitude
$1/(d_A^2{-}1)(d_B^2{-}1)$, with the inclusion-exclusion signs
of the harmonic-number combination.  {Expanding the rational partial fractions as geometric series in inverse powers of $\alpha_i$ (which scale with $N$) and integrating term-by-term against the Bose--Einstein kernel}, each {asymptotic Taylor} coefficient {inherits} this common factor, generating the Bernoulli
series~(\ref{eq:bernoulli}).  Conversely, applying Binet's
formula~(\ref{eq:binet}) to each partial fraction reconstructs the
digamma representation~(\ref{eq:Idiag}) identically: the integral
and digamma forms are not merely asymptotically equivalent but
exactly equal.

The factorisation
$\langle I\rangle = (d_A^2{-}1)(d_B^2{-}1)\,\mathcal{G}$ with
$\mathcal{G} = 1/(2N)-2\mathcal{J}$ is the central result.  It is
stronger than the order-by-order Bernoulli
statement~(\ref{eq:bernoulli}): the vanishing of mutual information
whenever either subsystem is trivial ($d=1$) is a property of the
\emph{exact} answer, not merely of each asymptotic coefficient.  The
``bipartite Casimir count''
$\dim[\mathfrak{su}(d_A)]\cdot\dim[\mathfrak{su}(d_B)]$ governs the
mutual information non-perturbatively.

Since $d^{2k}-1 = (d^2{-}1)(d^{2k-2}+d^{2k-4}+\cdots+1)$ for
any $k\ge 1$, the overall factor $(d_A^2{-}1)(d_B^2{-}1)$ divides
every term in the Bernoulli series~(\ref{eq:bernoulli}), so the
all-orders factorisation is already visible at the asymptotic level.
The integral representation~(\ref{eq:closed}) promotes this to the
exact, non-perturbative statement.

As a numerical check, for $d_A=2$, $d_B=3$, $d_E=7$ ($N=42$) the
exact Page formula gives
$\langle I\rangle = 0.28458\ldots$; evaluating
$(d_A^2{-}1)(d_B^2{-}1)\,[1/(2N)-2\mathcal{J}]$ by numerical
quadrature reproduces this to $15$ significant digits.

\emph{Manifest upper bound.}---The integrand of $\mathcal{J}$ is
not pointwise positive: the rational factor
$R(u) = u(C^2{-}u^4)/\prod_i(u^2{+}\alpha_i^2)$ changes sign at
$u = \sqrt{C}$.  However, $R$ is odd under the scale inversion
$u\to C/u$, meaning $R(C/u)\cdot C/u^2 = -R(u)$, while the Jacobian
$d(C/u) = -(C/u^2)\,du$ provides a compensating sign, so
$R(C/u)\,d(C/u) = R(u)\,du$.  Folding the integration domain at
$u = \sqrt{C}$ gives
\begin{equation}
\mathcal{J} = \int_0^{\sqrt{C}}\!R(u)
\biggl[\frac{1}{e^{2\pi ud_E}{-}1}
      -\frac{1}{e^{2\pi(C/u)d_E}{-}1}\biggr]du.
\label{eq:folded}
\end{equation}
On $(0,\sqrt{C})$ we have $u < C/u$; the Bose--Einstein function
$f(x) = 1/(e^{2\pi xd_E}{-}1)$ is strictly decreasing, so
$f(u) > f(C/u)$; and $R(u) > 0$ since $u^4 < C^2$.  The integrand
is therefore pointwise positive, proving $\mathcal{J}>0$ and hence
the strict bound
\begin{equation}
\langle I(A{:}B)\rangle
< \frac{(d_A^2{-}1)(d_B^2{-}1)}{2N}\,,
\label{eq:bound}
\end{equation}
with the deficit controlled by the single explicit integral
$\mathcal{J}$.  Neither the digamma representation nor the divergent
Bernoulli series makes this bound manifest.

\emph{Discussion.}---The value of $\langle I(A{:}B)\rangle$ is
fully determined by Page's formula and has been computable since the
proofs of Refs.~\cite{Foong1994,SanchezRuiz1995,Sen1996}.  Sen's
Bernoulli expansion~\cite{Sen1996} of individual subsystem entropies
is also well known.  The new content of Eq.~(\ref{eq:closed}) is
threefold.  First, it provides a \emph{non-perturbative completion}:
the Bernoulli series~(\ref{eq:bernoulli}) diverges, so no finite
truncation converges to the exact answer; Eq.~(\ref{eq:closed}) is the
Borel sum of that divergent series, expressed as a single convergent
integral.  Second, it makes the \emph{exact} factorisation by
$(d_A^2{-}1)(d_B^2{-}1)$ visible, establishing that the bipartite
Casimir count governs the mutual information non-perturbatively.  The
partial-fraction decomposition~(\ref{eq:pf}) shows that this
factorisation has a simple structural origin: all four poles of the
kernel have equal-magnitude residues.  Third, the strict upper
bound~(\ref{eq:bound}) becomes manifest through the scale-inversion
symmetry, without appeal to external inequalities.

The derivation rests on the diagonal/eigenvalue decomposition of
Page's entropy~(\ref{eq:page}).  The diagonal contribution,
the Dirichlet entropy of the simplex probabilities, produces the
digamma combination~(\ref{eq:Idiag}), while the Schur-majorisation
eigenvalue correction produces the rational
terms~(\ref{eq:Deltaev}) carrying the $\mathfrak{su}$ and Cartan
dimension counts.  The equal per-generator Bloch
variance~(\ref{eq:equalvar}) confirms that this separation is a
property of the nonlinear entropy functional, not of the state
geometry: the purity, being quadratic in $\rho$, is blind to the
simplex boundary.  It is only the concavity of $-x\ln x$ that
maps the $(m{-}1)$-dimensional simplex constraint into the entropy
reduction, producing the Cartan count in the eigenvalue penalty.

We also note a connection to entanglement detection.  The leading
asymptotic value $(d_A^2{-}1)(d_B^2{-}1)/(2N)$ is commonly used as a
benchmark for the ``typical'' mutual information.  The
bound~(\ref{eq:bound}) shows that this benchmark is a strict
\emph{overestimate}, with the fractional correction controlled by
$\mathcal{J}/[1/(2N)]$.  For small environments ($d_E$ not much
larger than $d_Ad_B$), this correction can be significant: for
$d_A=d_B=2$ and $d_E=4$ ($N=16$), the exact mutual information is
approximately {$1\%$} below the leading-order value.

The appearance of the Bose--Einstein kernel $1/(e^{2\pi t}{-}1)$ and
the Riemann zeta values $\zeta(1{-}2k)$ are mathematical in origin
(they follow from Binet's representation of $\psi$), but they place
the finite-size entanglement corrections in structural parallel with
one-loop Casimir calculations.  Whether this parallel admits a
physical interpretation, through replica field theory, holographic
entanglement, or the information geometry of quantum
states~\cite{Bengtsson2006}, is an open question.

All results require $d_Ad_B\le d_E$.  When this condition is
violated, the rational correction for $S(AB)$ in Page's formula
changes from $-(d_Ad_B{-}1)/(2d_E)$ to $-(d_E{-}1)/(2d_Ad_B)$
and the symmetric factorisation breaks: for $d_A=3$, $d_B=4$,
$d_E=2$, the exact value is $\langle I\rangle\approx 1.378$ while
the factorised form gives $\approx 2.483$.  However, the
diagonal/eigenvalue split~(\ref{eq:Isplit}) remains well-defined,
with only the rational piece requiring modification; the harmonic-number combination continues to equal
$\langle I_{\mathrm{diag}}\rangle$ regardless of dimension
ordering.  Extension to R\'enyi mutual informations would require
Weingarten calculus~\cite{Collins2006} for higher moments and does
not currently admit a comparable Borel-summable representation;
whether such representations exist is a natural direction for
future work.

Finally, we note a structural connection to typicality bounds.
Conventional concentration-of-measure arguments (e.g.\ L\'evy's
lemma~\cite{Popescu2006}) bound the probability that a subsystem
deviates from the microcanonical average but do not resolve the
internal structure of the deviation.  The decomposition~(\ref{eq:Isplit})
provides this resolution for the mutual information: the classical
(diagonal) and quantum (eigenvalue) contributions are separated
exactly, with the former captured by a Dirichlet entropy and the
latter by a Schur-majorisation correction.  The closed
form~(\ref{eq:closed}) then packages both into a single expression
whose convergent integral representation may prove useful in
finite-size analyses of quantum thermodynamic protocols where the
Bernoulli series is insufficiently accurate.

\bigskip



\end{document}